\documentclass[runningheads]{llncs}
\usepackage{graphicx}
\usepackage{times}

\usepackage{soul}
\usepackage{url}
\usepackage[hidelinks]{hyperref}
\usepackage[utf8]{inputenc}
\usepackage[small]{caption}
\usepackage{amsmath}
\usepackage{booktabs}
\usepackage{multirow}
\usepackage{algorithm}
\usepackage{algorithmic}
\renewcommand{\ldots}{\!...}
\usepackage[colorinlistoftodos, textwidth=19mm, textsize=footnotesize, shadow]{todonotes}

\usepackage{xcolor}

\begin{document}
\title{The Parking Problem: A Game-Theoretic Solution}

\author{Giuseppe Calise \and Aniello Murano \and
Silvia Stranieri}
\authorrunning{G. Calise and A. Murano and S. Stranieri}

\institute{University of Naples, Federico II, Italy\\
\email{calisegiuseppe@outlook.com,\{aniello.murano,silvia.stranieri\}@unina.it}}
\maketitle            
\begin{abstract}
In this paper, we propose a game-theoretic solution to the parking problem, by exploiting a strategic-reasoning approach for multi-agent systems. Precisely, cars are modeled by agents interacting among them in a multi-player game setting, whose aim is to get a free slot parking-place satisfying their own constraints. The overall assignment is then given as a Nash equilibrium solution.
We come up with an algorithm (and its implementation in a tool) that works in quadratic time. We give evidence of the benefits of our approach by running our tool on a large hospital parking space. 
\end{abstract}

\section{Introduction}

With the fast development of economy and city modernization, traffic congestion and parking have become serious social problems. Studies conducted in big cities report that daily, on average, drivers take more than eight minutes to park, causing the $30\%$ of traffic~\cite{ayala2011parking,shoup2005high}. 
Such statistics raise several side effects, among which a high fuel consumption, high $CO_2$ emissions, but also a stressful lifestyle for drivers. 
The growth of Artificial Intelligence applications to automotive 
is constantly increasing the request for smart solutions to parking. 
This research field is well identified as \emph{smart parking} (see~\cite{lin2017survey}). The competitive nature of the parking process, during which the drivers compete in order to get an available parking slot for their cars, is the inspiration of this work. Indeed, by exploiting basic settings of the strategic reasoning for multi-agent systems, we model the parking process as a competitive multi-player game in which each car is an agent interacting with all the other ones, with the  ultimate goal of getting an available slot that satisfies its own constraints. The \emph{parking problem} we face concerns parking as many cars as possible, while satisfying their requirements. 

A multi-agent system is made of autonomous entities, with distributed 
information, computational capabilities, and possibly diverging interests. 
These kind of systems have been exploited in several fields: 
electronic~\cite{song2021cooperative} and industrial 
process control~\cite{bakliwal2018multi}, economy~\cite{pal2018multi},
home automation~\cite{zouai2017smart}, 
open system verification~\cite{alur2002alternating}, to name a few. The way autonomous agents can interact with each other can be classified into two categories: competitive and cooperative. In the former case, there is no a-priori agreement among agents, as they try to maximize their own 
objective, no matter what the objectives of the other agents are. 
In the latter case, the agents coordinate among them in order to get the best outcome possible for everyone~\cite{wooldridge2009introduction}. 

{\bf Our contribution.} We address the parking problem by means of a multi-agent strategic-reasoning approach. Specifically, we model the parking problem as a multi-agent game where cars are competitive agents, moving concurrently and under perfect information. We assume that each agent comes with a desired time-limit to accomplish the car parking. Also, for each slot, we have a time needed to be reached from each entrance. Then, for an agent, the choices for a slot are strategies whose payoff reflects the maximum time he consumes to park his own car (or the fact that he cannot park at all). Solving the parking problem corresponds to finding a solution in such a multi-agent game that minimizes all agents' payoffs. We find such a solution by means of Nash equilibrium and prove its effectiveness on real scenarios. We recall that in a game a Nash equilibrium is reached when each player does not have any incentive to unilaterally change his strategy. 

The contribution of this work is twofold. From one side, we come up with an effective multi-agent game model for the parking problem we consider. From the other side, we provide an algorithm (and its implementation in a tool), working in quadratic time, that allows a fair allocation of the parking slots by satisfying a Nash equilibrium. As we prove later on practical scenarios, this is a valuable compromise with respect to an optimal, but exponential, brute-force solution that would check all possible distributions of cars over available slots. 
Also, as expected, our solution is better than any greedy FIFO approach. Indeed, consider a scenario in which there are three 
vehicles, $V_1$, $V_2$, and $V_3$, looking for a parking, and   
three slots available $A$, $B$, and $C$. Assume now that $V_1$, 
$V_2$, and $V_3$ have up to $7$, $5$, and $3$ minutes to accomplish the parking, respectively. Also, assume that slots $A$, $B$, and $C$ 
require $2$, $3$, and $5$ minutes to be reached, respectively. Assume 
now that $V_1$ picks $A$ and $V_2$ picks $B$; then, 
$V_3$ would not have enough time to reach the remaining slot $C$.
Contrarily, a solution that allows parking all vehicles by accommodating their requirements is to assign $V_1$, $V_2$, and $V_3$ to $C$, $B$, and $A$, respectively. 
This is exactly what our algorithm would return as a solution.

Note that the multi-agent game model we set up can also admit more than one Nash equilibrium. In game theory, in general, this is problematic as the players do not know which one to choose. 
In our setting, however, this is not a problem as it is the system that chooses just one equilibrium and all cars will be instructed to behave accordingly.

\section{Related Works}

Smart parking solutions literature is very reach and diversified. In \cite{lin2017survey}, the authors provide a large survey on smart parking modeling, solutions, and technologies as well as identify challenges and open issues.
Algorithmic solutions have been also proposed in the VANET research 
field, see for example \cite{senapati2020automatic,rad2017smart,safi2018svps,balzano2019acop,DBLP:conf/euspn/BalzanoMV16,DBLP:journals/jhsn/BalzanoMS17}. Less common is the use of game-theoretic approaches to address the parking problem. An exception is \cite{kokolaki2013efficiency}, 
which is probably the closest to us, indeed the authors also
propose a parking solution based on the Nash equilibrium. However, differently from us, they provide a numerical solution (rather than an algorithm or a tool), and, more importantly, they consider a scenario with both private and public parking slots, and the drivers' payoffs strongly rely on such a topology.
Smart parking mechanisms based on a multi-agent game setting have been also proposed in the literature. In \cite{malecki2018computer}, drivers' behavior is simulated by modeling the environment on the basis of cellular automata. In \cite{belkhala2019smart} the model is based on the interaction between the user (driver) and the administrator, but focusing more on the architecture rather than the model setting and the strategic reasoning. Similarly, \cite{jioudi2019parking} provides an E-parking 
system, based on multi-agent systems aimed to optimize several users' 
preferences. In \cite{okoso2019multi}, the authors manage the 
parking problem with a cooperative multi-agent system, by relying on 
a priority mechanism. In \cite{pereda2020competing}, the authors also focus 
on an equilibrium notion, but they study the Rosenthal equilibrium rather 
than the Nash one, which describes a probabilistic choice model. 
Finally, \cite{lu2021equilibrium} also considers the concept of Nash equilibrium applied to cars, but it is used to talk about 
traffic rather than parking.

In this work, we not only design the parking process as a game among agents playing competitively, but also study the use of the Nash equilibrium as a solution. To the best of our knowledge, this is the first work addressing the parking problem via multi-player game, whose solution is given algorithmically by solving a Nash equilibrium.

\section{A Real Scenario} \label{scenario}

\begin{figure}[h]
\vspace{-0.5em} 
\centering
  \includegraphics[width=0.80\linewidth]{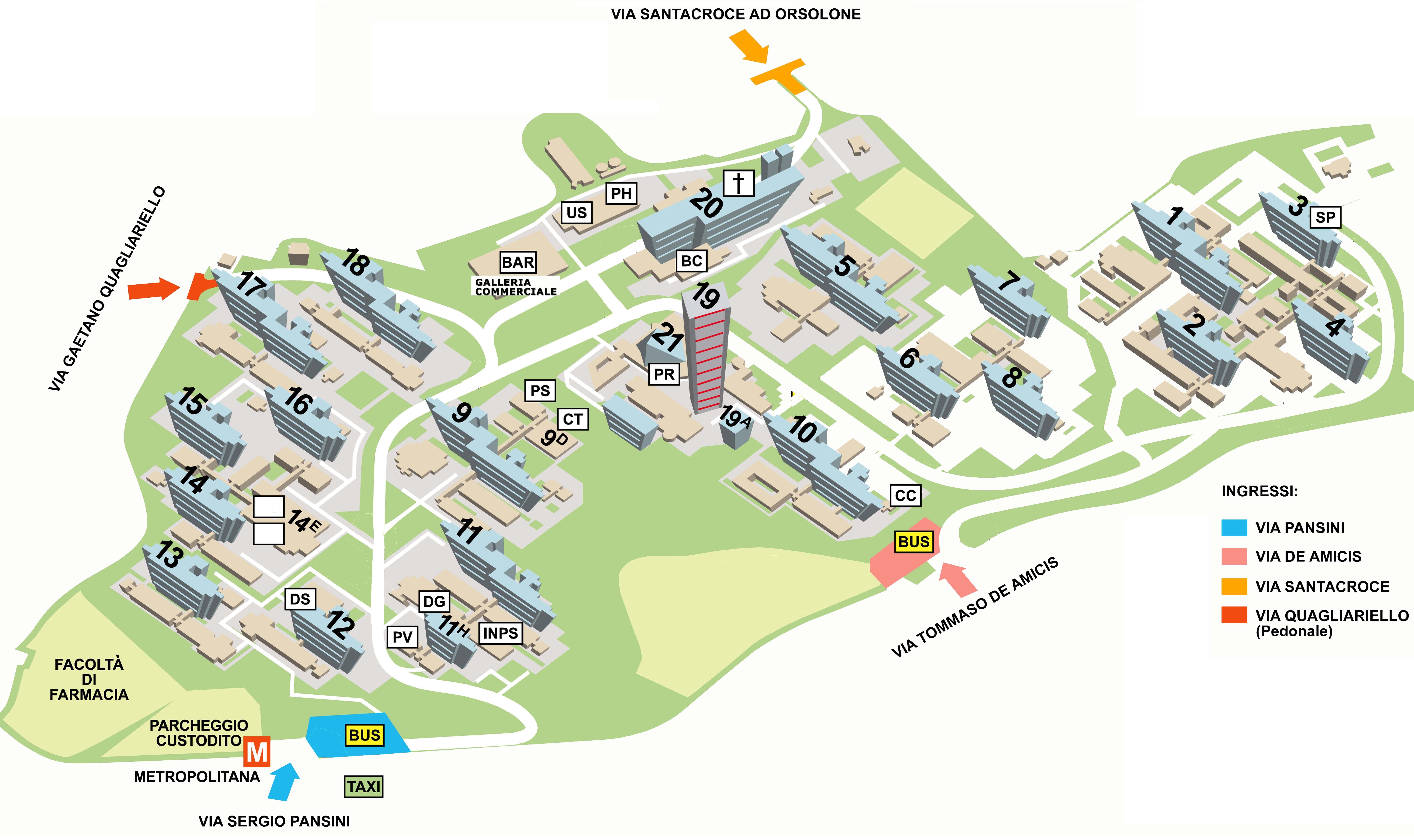}
\vspace{-0.5em}
\caption{Graphical representation of the A.O.U. Federico II}
\vspace{-0.5em}
\label{fig:AOU}
\end{figure}

As case study, we have focused on the parking area of the \textit{Federico II Hospital Company} in Naples, one of the biggest and most specialized hospital in the South of Italy, whose construction goes back to the early Sixties. The Hospital, as it is depicted in Figure~\ref{fig:AOU}, is made of $21$ building blocks, distributed over $440000 m^2$, and provides in total one thousand of beds for ordinary recovery and two hundreds of beds for day-hospital use. The parking space, having $2684$ slots in total, consists of $21$ independent areas, and is mainly used by patients and, in turn, by the $3400$ employees (doctors, nurses, technicians, administrators, etc.). The hospital has four guarded gates, one of which is for pedestrian. The car gates are preceded by a road where cars line up for the necessary checks. On average, it is estimated that there are $4600$ car accesses per day. There is no policy about the allocation of the parking places and, except for few reserved ones, each driver chooses by its own the slot. This disorganized solution produces a huge traffic congestion, bottlenecks at the entrance, and an unbalanced distribution of the cars over the parking area. More importantly, it does not take into account the specific constraints and some physical limitations of the users, such as walking issues or urgency. In the most crowded hours, on average, the drivers spend more than 20  minutes to find a parking slot or, even worst, they leave the parking area by missing available slots. 

In order to efficiently apply our tool, we assume that the list of available slots in every area of the hospital is known at runtime. Also, we make use of all information the car passengers have to communicate to the hospital before entering, and in particular their logistics. Finally, we assume that the drivers will be followed while driving inside the parking area, by means of tracking devices (GPS, smartphone, videocameras, etc.).  

Having all this information at its disposal, the tool works as follows: it takes all cars in queue on the roads in front the car gates, as well as all the specific needs and constraints of their occupants. Then, it processes the data and following the algorithm described in the sequel, it opportunely associates the available slots to the cars. In particular, the tool will access both the Employers Data Center and the Online Booking Center of the hospital and, thanks to the latter, the tool will know which kind of services the patients need, date and time of their appointments, possible walking limitations and handicaps, etc. Note that the tool operates in stages, processing one bunch of cars at the time, as they are in queue. Someone may criticize this solution and propose an offline allocation instead. We decide not to follow this solution for two main reasons: first, the hospital is highly dynamic in slot requests and, more importantly, slots are very limited in numbers, so it is better to allocate slots only when cars show up.

\section{Parking Game Structure}
In this section we introduce the \textit{Parking Game Structure} model, (\emph{PGS}, for short), that is the game model we will use along our algorithm to solve the parking problem we address. 
The model definition takes inspiration from the scenario described in Section \ref{scenario}. Thus, in a PGS, the players are cars with their needs and constraints. Also, the PGS takes into account all the specification about the slots, in particular their location, their availability, the time they require to be reached from each entrance, and so on.

Formally the \emph{Parking Game Structure} is defined as follows:

\begin{definition}\mbox{[Parking Game Structure]\label{game}
    The Parking Game Structure(PGS) is a tuple:}
    \begin{equation}
        \mathcal{G} = (Agt, S, G, g, F, T, R)
    \end{equation}
\vspace{-0.3em}
where:
\vspace{-0.3em}
    \begin{itemize}
    \item $Agt = \{a_1, \ldots, a_n \}$ is a set of \emph{agents}, i.e., the cars,
    \item $S = \{s_1, \ldots, s_m \}$ is a set of parking \emph{slots},
    \item \mbox{$F = \{f_1, \ldots, f_n \mid f_i \in [0,1]\}$ is a set of \emph{resilience} values,}
    \item $G = \{g_1, \ldots, g_l \}$ is a set of \emph{gates},
    \item $g: Agt \rightarrow G$ is a function associating agents to gates,
    \item $AT = \{t_1, \ldots, t_n\}$ is a set of \emph{agent-time} values, where $t_i$ is the time limit the car $a_i$ has for parking, 
    \item $RT = \{r_{(1,1)}, \ldots, r_{(m,l)}\}$ is a set of \emph{reaching-time} values, where $r_{(i,j)}$ is the time needed to reach the parking slot $s_i$ from gate $g_j$.
    \end{itemize}
\end{definition}

Regarding the set of resilience indexes $F$, note that 
each $f_i$ is associated with agent $a_i$ and it has a twofold use: first, it imposes an order among agents; second, it affects the final pre-emption order. This will be more clear below. For simplicity, we assume that all the resilience indexes are different, i.e., $f_i \neq f_j$ for every $1 \leq i < j \leq n$. The indexes in $F$ can be set manually as input, however we report that, for the case study we have introduced in Section \ref{scenario}, the values have been obtained automatically by processing the information coming from the Employers Data Center and the Online Booking Center of the hospital; in particular, for the patients, the resilience index represents their movement ability, therefore, the lower the rate, the more favored the patient. 

A \textit{strategy} for an agent $a_i$ consists of choosing a slot $s_j \in S$. Formally it is a function $Str: Agt \rightarrow S$. A \textit{strategy profile} is an $n$-uple $\overline{s}=(\overline{s_1}, \ldots, \overline{s_n})$ of strategies, one for each player. Formally, in $\overline{s}$, for each $i$, we have $Str(a_i)=\overline{s_i}$. It is worth noting that it may happen that two or more players choose the same strategy. Next we define the \emph{costs associated} to $\overline{s}$ as a tuple of costs $\overline{c}=(\overline{c_1}, 
\ldots, \overline{c_n})$. Then, a \textit{payoff} $\pi$ of a strategy $\overline{s}$ is defined as a sum of all such $\overline{c_i}$, i.e., $\pi(\overline{s})=\sum_i \overline{c_i}$, and by $\pi_i$ we denote the $i-$th cost value of that tuple.

\begin{definition}
Let $a_i\in Agt$ be an agent with $g(a_i)=h$ and $\overline{s}=(\overline{s_1}, \ldots, \overline{s_n})$ be a strategy profile, with $\overline{s_i}=s_j$ for an $s_j \in S$. We define the \emph{costs associated} to $\overline{s}$ as the tuple $\overline{c}=(\overline{c_1}, 
\ldots, \overline{c_n})$ where each $\overline{c_i}=$ 
\vspace{-0.5em}

\begin{eqnarray} \label{eq:cost}
    \begin{cases}
    
    f_i \cdot (t_i - r_{(j,h)}) & \hspace{-0.1em} \textit{if 1. } (t_i - r_{(j,h)}) \geq 0 
    \textit{, and}
        \\ & 
        \hspace{0.7em} \textit{2. there is no } a_{k \neq i} \textit{ such that }
        \\ & 
        \hspace{0.7em} f_k < f_i \textit{, } 
        \overline{s_k} = \overline{s_i} \textit{, }
        \textit{ and } 
        \\ &
        \hspace{0.7em} (t_k - r_{(j,p)}) \geq 0 
        \textit{, with } g(a_k)=p
        \\
        \infty & \textit{otherwise}
    \end{cases}
\end{eqnarray}

\end{definition}

In words, the value $\overline{c_i}$ is a finite value if the agent $a_i$ has enough time to reach the parking slot $s_j$ and such a slot has not been taken from any other agent $a_k$ with a lower resilience (i. e., $f_k < f_i$). Then, the value, when it is finite, reflects how much time it is left to the agent after he has reached the assigned slot (with respect the total time he has at his disposal). 
Conversely, the infinity value corresponds to the worst possible outcome for the agent $a_i$, which reflects the fact that he cannot park at the slot $s_j$. 
At this point, it should be intuitive that the problem of looking for an optimal strategy profile $\overline{s}$ can be reduced to the problem of minimize\footnote{Note that the minimization guarantees that the best slots are kept for future use, so to focus on the continue allocation process rather than the single stage. At any rate, one can use maximization (in a finite domain) without affecting the rest of the algorithm.} the corresponding vector of associated costs $\overline{c}$.

Unfortunately, this is in general not an easy task. In particular, a brute-force algorithm checking all the possible strategy profiles is unfeasible as it requires exponential-time. Conversely, we suggest adopting a Nash equilibrium solution that provides, by definition, a satisfactory solution and, along with our setting, it just requires quadratic-time. In the sequel we are going to present such a solution. Also, we present a solution based on a greedy behavior of the players, which reflects the current behaviour of drivers at the parking of the hospital described in Section \ref{scenario}: each car takes the first available parking slot which satisfies its needs. By means of a toy example, we show that the solution based on the Nash equilibrium over-perform the one based on the greedy behaviour of the players.

\section{The Parking Slot Selection Game}
In this section, we first provide a toy example, then we introduce the \textit{Parking Slot Selection Game} (PSSG, for short) and propose a solution by means of a Nash equilibrium calculation. We also comment on the greedy approach and compare it with our solution. For a matter of presentation, we will recall the notion of Nash equilibrium.

Before proceeding, it is worth noting that at each instance of the  game we consider, each car can enter the parking space through just one entrance. This means that we can get rid of $g$ and $G$ when dealing with a PGS, as well as the second index of the reaching-time values in $RT$. This also allows us using a simplified version of the definition of costs associated with strategy profiles. In other words, while the set $RT$ provides $m*l$ possible reaching values in general as stated in Definition \ref{game} (with $m$ the number of slots and $l$ the number of gates), each instance of the game just requires dealing with $RT$ as a vector of $m$ values, i.e., $RT = \{r_1,\ldots,r_m\}$, where each $r_i$ represents the time needed to reach the parking slot $s_i$ from the physical gate through which the car is entering. When providing our solution to PSSG in Algorithm \ref{algo}, we strongly rely on this observation, which leads to a natural reformulation of the model right after the vehicles are associated to the gates. 
Notably, we prefer to keep our PSG model as general as possible in order to accommodate other questions that require dealing with not \emph{a priori} fixed entrances associated to cars. For example, it may be useful when devising an algorithm that also suggests in advance to a driver the gate to take. This, however, is not the target of this paper.

\subsection{A Toy Example} \label{sub:example}
Let us consider a parking place with $3$ slots available and $3$ cars aiming at parking. 
\begin{figure}[h]
\vspace{-0.5em}
  \centering
  \includegraphics[width=0.50\linewidth]{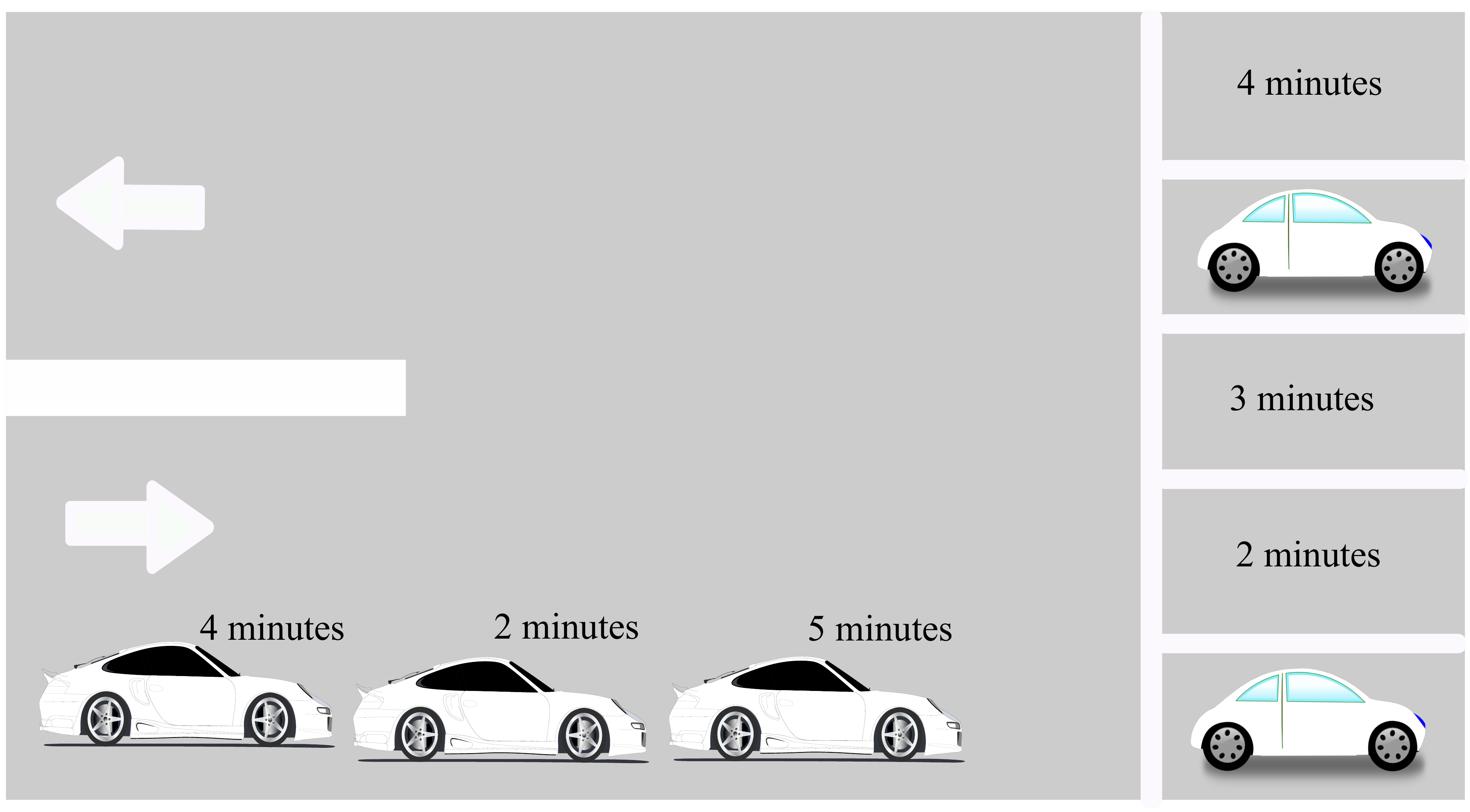}
  \vspace{-0.5em}
  \caption{3-players-3-slots Game.}
  \vspace{-0.5em}
  \label{fig:3p3s}
\end{figure}
Let us suppose that the first, the second, and the third car have respectively $5$, $2$, and $4$ minutes available to park and that, as associated resilience they have $0.5$, $0.1$, and $0.009$, respectively.
Also, suppose that the first, the second, and the third slot require $2$, $3$, and $4$ minutes to be reached, respectively. We call such a game the $3$-players-$3$-slots game and it is reported in Figure \ref{fig:3p3s}.

\subsection{A Greedy Solution}
When a car is approaching to the parking, a greedy solution is to occupy the first slot it can get. This approach leaves to the car a free will to 
park in the slot that best fits its constraints, without paying 
attention to the other cars requirements. This easy-to-design solution may 
lead to a non-optimal vehicles allocation, as it may leave out some cars 
(not able to park), as the remaining slots may not satisfy their requirements. 

To give an example, let us consider the scenario described in Section 
\ref{sub:example}. In this situation, the first car would choose the 
closest slot (the one that requires $2$ minutes to be reached). Then, the second car would not be able to park, because all the remaining free 
slots are too expensive in terms of time. 

For this reason, we have looked for a better solution that would exploit the car parking potentialities at the best by means of a smart distribution of slots among cars, and that would be computationally easy to be calculated on the fly.

\subsection{Nash Equilibrium Based Solution}
In game theory, a well-conceived solution concept that ensures a robust 
form of satisfaction among players is \textit{Nash equilibrium}. 
This concept was deeply investigated and well formalized by John Nash in the 
fifties, both under pure and mixed strategies (see \cite{van1991stability} 
for more details). In the basic definition, we say that in a multi-player game, all 
players, moving concurrently, reach a Nash equilibrium if none of them has 
the incentive to unilaterally deviate from that equilibrium. By casting this 
in our parking scenario, we try to reach a situation in which all drivers are 
associated to parking slots, by means of an equilibrium over their constraints. 
In other words, our goal is to provide a strategic profile (parking slot 
assignment) in which no player wants to change his slot unless some other 
players want to change theirs.

Following the model definition given in Definition \ref{game} and the 
observations made above, we formally introduce the \textit{Parking Slot 
Selection Game} we address, as follows.

\begin{definition}[Parking Slot Selection Game] \label{def}
The Parking Slot Selection Game (PSSG) has an input and an output defined as follows:
\begin{itemize}
    \item Input: a PGS $\mathcal{G}$, as given in Definition \ref{game}.
    \item Output: a strategic profile $(\overline{s}_1^*, \ldots, \overline{s}_n^*)$ providing a Nash equilibrium for $\mathcal{G}$. A strategy profile $(\overline{s}_1^*, \ldots, \overline{s}_n^*)$ is a Nash equilibrium if and only if $\forall \, \overline{s}_1, \ldots, \overline{s}_n \in S$ it holds that

        $\pi_1(\overline{s}_1^*, \ldots, \overline{s}_n^*) \leq \pi_1(\overline{s_1}, \overline{s}_2^* \ldots, \overline{s}_n^*)$, 
        $\pi_2(\overline{s}_1^*, \ldots, \overline{s}_n^*) \leq \pi_1(\overline{s}_1^*, \overline{s_2} \ldots, \overline{s}_n^*), \dots ,$

        $\pi_n(\overline{s}_1^*, \ldots, \overline{s}_n^*) \leq \pi_1(\overline{s}_1^*, \overline{s}_2^* \ldots, \overline{s_n}).$

\end{itemize}

\end{definition}

In words, the PSSG consists in looking for a strategy profile that, with respect to the associated costs, no player has an incentive to unilaterally change his choice. 

Similarly to the PSGG, one can define the \emph{Greedy Parking Game} (\emph{GPG}, for short). To give some details, first assume that in an GPG players are ordered, then the strategy profile $(\overline{s}_1^*, \ldots, \overline{s}_n^*)$ is such that for each agent $a_i$, it holds that $\overline{s}_i^*$ is the best choice (in terms of minutes to reach it) over $S \setminus \{\overline{s}_1^*, \ldots, \overline{s}_{i-1}^*\}$.

\subsection{A Solution to the 3-players-3-slots Game}
Let us consider again the 3-players-3-slots game described in Section 
\ref{sub:example}. We now show a solution based on the satisfaction of a 
Nash equilibrium. As we will see in a while, such a solution allows accommodating all cars, while satisfying all their constraints, contrarily 
to what we have seen with the greedy solution. Later, we will show that 
this is true in general and not just for the case of our specific example. 
Let us formally describe the 3-players-3-slots example by means of a PGS $\mathcal{G}_3$ whose components are defined as follows:

\begin{itemize}
    \item $Agt = \{car_1, car_2, car_3\}$ is the set of cars,
    \item $S = \{{slot}_1, {slot}_2, {slot}_3\}$ is the set of parking slots,
    \item $AT = \{5, 2, 4\}$ is the set of time-values $car_1$, $car_2$, and $car_3$ have at their disposal, respectively,
    \item $RT = \{2, 3, 4\}$ is the set of times needed to reach the slots ${slot}_1$, ${slot}_2$, and ${slot}_3$, respectively,
    \item $F = \{0.5, 0.1, 0.009\}$ is the set of cars resilient values, 
    \item The cost function is reported in Table \ref{tab:tablecost}, in the last three rows. For instance, the triple $(\infty, \infty, 0.018)$ represents the case in which all cars decide to park in the same slot ${slot}_1$; so, $car_3$, which has the lowest resilience value, gets it at a cost of $0.018$ (i.e., $(4 - 2) \cdot 0.009$), while the other cars leave the process incomplete, as they get $\infty$.
\end{itemize}

\begin{table*}[ht!]
\tiny
    \centering
    \resizebox{12.5cm}{1.2cm}{%
    \begin{tabular}{cc|c|c|c|c|c|c|c|c|c|}
        \cline{3-11}
        &  & \multicolumn{9}{c|}{$car_3$} \\ 
        \cline{3-11}
        &  & \multicolumn{3}{c|}{${slot}_1$} & \multicolumn{3}{c|}{${slot}_2$} & \multicolumn{3}{c|}{${slot}_3$} \\ 
        \cline{3-11}
        &  & \multicolumn{3}{c|}{$car_2$} & \multicolumn{3}{c|}{$car_2$} & \multicolumn{3}{c|}{$car_2$} \\ 
        \cline{3-11}
        &  & ${slot}_1$ & ${slot}_2$ & ${slot}_3$ & ${slot}_1$ & ${slot}_2$ & ${slot}_3$ & ${slot}_1$ & ${slot}_2$ & ${slot}_3$ \\
        \hline
        \multicolumn{1}{|c|}{\multirow{3}{*}{$car_1$}} & ${slot}_1$ & $\infty,\infty,0.018$ & $\infty,\infty,0.018$ & $\infty,\infty,0.018$ & $\infty,0,0.009$ & $1.5,\infty,0.009$ & $1.5,\infty,0.009$ & $\infty,0,0$ & $1.5,\infty,0$ & $1.5,\infty,0$ \\ \cline{2-11}
        
        \multicolumn{1}{|c|}{} & ${slot}_2$ & $1,\infty,0.018$ & $1,\infty,0.018$ & $1,\infty,0.018$ & $\infty,0,0.009$ & $\infty,\infty,0.009$ & $\infty,\infty,0.009$ & $\textbf{1,0,0}$ & $1,\infty,0$ & $1,\infty,0$ \\ \cline{2-11}
        
        \multicolumn{1}{|c|}{} & ${slot}_3$ & $0.5,\infty,0.018$ & $0.5,\infty,0.018$ & $0.5,\infty,0.018$ & $0.5,0,0.009$ & $0.5,\infty,0.009$ & $0.5,\infty,0.009$ & $\infty,0,0$ & $\infty,\infty,0$ & $\infty,\infty,0$ \\ \hline
    \end{tabular}%
    }
    \caption{Cost function values for 3-drivers-3-slots instance of the game.}\label{tab:tablecost}
\end{table*}

By a matter of calculation, one can check that there exists only one Nash equilibrium, which corresponds to $\overline{s}= ( {slot}_2, {slot}_1, {slot}_3 )$, with $\overline{c} = (1, 0, 0)$ (in bold in Table \ref{tab:tablecost}), and $\pi(\overline{s})=1$.

\subsection{A Solution to the Parking Slot 
Selection Game}

In this section, we introduce the algorithm for the solution to the problem described in Definition \ref{game}. We first provide the pseudo-code in 
Algorithm \ref{algo}, then we describe how it works and report on its time complexity.

\begin{algorithm}[h]\label{algo}
\caption{Algorithm for the solution of the PSSG.}
\label{algo}
\textbf{Input}: Queue of ready vehicles\\
\textbf{Output}: Slot allocation
\begin{algorithmic}[1] 
\WHILE{$carQueue \neq null$}
	\STATE $actualCar = priorityCar(carQueue)$.
	\STATE $outcome = \infty$.
	\FOR{$slot \in setAvailableSlots$}
		\STATE $po = c(actualCar, slot)$.
		\IF {$po \geq 0 \And po < outcome$}
			\STATE $outcome = po$.
			\STATE $assignSlot(actualCar, slot)$
			\STATE $setNotAvailable(slot)$.
		\ENDIF
	\ENDFOR
\ENDWHILE
\end{algorithmic}
\end{algorithm}

With the first iteration, the car with the lowest resilience index, \textit{actualCar}, is selected from the queue, through the function \textit{priorityCar($\cdot$)}, which takes as input the set of cars and returns the one with the lowest resilience index respect to the others. The variable cost \textit{outcome} is associated an infinity value, the worst possible one. In the second iteration, the algorithm computes the costs resulting from the function \textit{c($\cdot$)}, which takes as input a car and a slot. The value of the \textit{outcome} is updated with the value of the best cost computed. Among the available slots, the one with the best result is assigned to the \textit{actualCar}. Once assigned, the slot is remove from the set of the available ones, with the function \textit{setNotAvailable($\cdot$)}.

\begin{theorem}[Correctness of Algorithm \ref{algo}]
Algorithm \ref{algo} computes the Nash equilibrium for the game.
\begin{proof}[Proof (Sketch)]
Proving that the algorithm provides a Nash equilibrium is quite trivial. Assume by contradiction that $\overline{s}=(\overline{s_1},..,\overline{s_n})$ is the solution provided from our algorithm and it is not a Nash equilibrium. Then, by definition of Nash equilibrium, there must exist an agent, let us say agent $a_i$, whose strategy $s_j$ is not the best, while fixed the strategies for the other players. Hence, there exists another strategy $s’_j$ for the agent $a_i$, such that the payoff of $s’_j$ is better than the one for $s_j$ (given the same strategies for the other players). But if such a strategy $s’_j$ exists, then it would be found at row $6$ of our algorithm, and it would be chosen as the final strategy for agent $a_i$. But this clearly contradicts the hypothesis that $\overline{s}=(\overline{s_1},..,\overline{s_n})$ is the solution provided.
\end{proof}
\end{theorem}

\begin{theorem}[Complexity of Algorithm \ref{algo}]
The complexity of Algorithm \ref{algo} is quadratic with respect to the number of 
agents involved in the game, in the worst case.

\begin{proof} 
Consider the worst possible scenario, i.e., no vehicle obtains a parking slot. Then, let us compute $\mathcal{C}(PSSG)$ as the complexity of the Parking Slot Selection Game. The proof proceeds by analyzing the complexity of the most expensive operations, 
from the inner ones to the outer ones. We use the notation $\mathcal{C}(r)$ to indicate the complexity of the code from the $r$-th row of the Algorithm \ref{algo}.

The function $assignSlot(Car,slot)$ performs simple assignments, with a constant 
complexity $\mathcal{C}(7)=O(1)$.

The inner loop does not perform any slot assignment, in the considered worst case, 
since none of them satisfies the constraints of the cars to be allocated. Hence, 
the inner loop is repeated $|S|$ times, where $S$ is the set of slots, 
according to Definition \ref{game}. Assuming that $|S|=m$, we can deduce that $\mathcal{C}(3)=\sum_{i=1}^{m} \mathcal{C}(7)=\sum_{i=1}^{m} O(1)=O(m)$.

The outer loop is performed as many times as the number of cars, i.e, the agents. As $|Agt|=n$ (Definition \ref{game}), we have $\mathcal{C}(1)=
\sum_{j=1}^{n} \mathcal{C}(3)=\sum_{j=1}^{n} O(k)=O(nm)$.

Assuming that, in the worst case, $n$ and $m$ are of the same order, we can conclude
that the total complexity is $\mathcal{C}(PSSG)=O(n^2)$.
\end{proof}
\end{theorem}


\section{Evaluation}\label{eval}

In this section, we compare the performances between executing the greedy solution to solve GPGs and Algorithm \ref{algo} to solve PSGGs. We have run $10$ times the two approaches on a growing number of cars and slots. All values and time-limit needed have been generated randomly. Results have been collected in Table \ref{comparison}. Each column represents a different execution of the two approaches 
with the corresponding input parameters, while the rows keep track of the two analyzed solutions. Each entry contains the number of cars that have been able to park successfully, over the total number of cars involved.
As one can observe, the Nash equilibrium based solution is never worse than the
greedy one. Moreover, by extending the experiment over $100$ and $200$ executions, our
approach is strictly better than the greedy one in the $89\%$ and $93\%$ of the cases respectively, and it allocates the same number of vehicles in the remaining ones. 

Since, by construction, a greater number of executions determines a greater 
number of cars, these experiments also prove the scalability of our algorithm,
which seems to behave well with high numbers. Such a scalability property will be explained in more details in the next section.

\begin{table*}[ht!]
    \centering
    \resizebox{0.8\textwidth}{!}{%
    \begin{tabular}{c|c|c|c|c|c|c|c|c|c|c|}
    \cline{2-11}
         & $E_1$ & $E_2$ & $E_3$ & $E_4$ & $E_5$ & $E_6$ & $E_7$ & $E_8$ & $E_9$ & $E_{10}$ \\ \cline{2-11}
         & 3 slots & 4 slots & 5 slots & 6 slots & 7 slots & 8 slots & 9 slots & 10 slots & 11 slots & 12 slots \\ 
         & 3 cars & 4 cars & 5 cars & 6 cars & 7 cars & 8 cars & 9 cars & 10 cars & 11 cars & 12 cars \\ \hline
        \multicolumn{1}{|c|}{PSSG}  & 3/3 & 3/4 & 5/5 & 6/6 & 7/7 & 8/8 & 7/9 & 8/10 & 9/11 & 12/12 \\ \hline
        \multicolumn{1}{|c|}{GPG} & 2/3 & 3/4 & 5/5 & 5/6 & 6/7 & 6/8 & 6/9 & 7/10 & 8/11 & 10/12 \\ \hline
    \end{tabular}%
    }
    \caption{Resulting vehicle allocations over 10 different simulations applying two solutions to the parking game: the Nash equilibrium based one, and the greedy one.}\label{comparison}
\end{table*}


We conclude this section by reporting some benchmarks regarding Algorithm \ref{algo}. Precisely, we have analyzed the behavior of the algorithm in the management of a growing number of cars waiting for a parking slot, with respect to a fixed number of parking slots. We have considered two scenarios and reported the corresponding benchmarks in Table \ref{tabletwo}. The first one considers $4600$ slots. Such a number is not picked at random, but it refers to the number of slots available inside the structure of our case study, including some private parking slots close by. The second one considers $20000$ slots. Also in this case, the number is not picked at random, but it refers to the number of available slots in the biggest parking space of the world (West Edmonton Mall in Canada).

\begin{table}[h]
\centering
\begin{tabular}{l|ccccccccc}
\multicolumn{1}{c|}{} & \multicolumn{9}{c}{seconds}                                                                                                                                                                                                                                \\ \hline
cars                  & 200                       & 400                       & 800                       & 1600                      & 3200                      & 6400                      & 12800                     & 25600                     & 51.200                      \\ \hline
4600 slots            & 0.001                     & 0.002                     & 0.004                     & 0.009                     & 0.027                     & 0.402                     & 1.389                     & 3.415                     & 10.165                     \\ \hline
20000 slots           & \multicolumn{1}{l}{0.003} & \multicolumn{1}{l}{0.006} & \multicolumn{1}{l}{0.013} & \multicolumn{1}{l}{0.026} & \multicolumn{1}{l}{0.060} & \multicolumn{1}{l}{0.150} & \multicolumn{1}{l}{0.430} & \multicolumn{1}{l}{5.687} & \multicolumn{1}{l}{23.597}
\end{tabular}
\caption{Results on 4000 and 20000 slots.}\label{tabletwo}
\end{table}

All tests have been executed on an Intel®Core™i5-7300HQ CPU processor of 2.50 GHz, with 8 Gb RAM capacity.

To show the scalability of our algorithm, we have considered a very large set of cars. The benchmarks show that our tool can be also used in other fields, with much higher numbers. For example, it can be used to accommodate people in a stadium, or, distribute people over hospitals, for example, for a massive vaccinations, as it is required nowadays for the Covid pandemic situation. 


\section{Conclusions and Future Works}\label{cfw}

The parking problem is one of the most challenging questions in the automotive research field. Inspired by the intrinsic interaction among cars that compete among them in order to get a parking slot complaining with their constraints, in this paper we have explored a game-based approach. Precisely, following a real case study, we have formally introduced (\textit{i}) a multi-player game structure model, (\textit{ii}) the problem, and (\textit{iii}) a solution algorithm working in quadratic time. The game model makes use of costs, reflecting the time ability of a car to park in a specific slot (modulo a resilience rate intrinsically associated to each car). The core of the algorithm is then based on a Nash equilibrium solution, which allows focusing not just on the best choice for a single car, but rather on one that guarantees a fair slot assignment among all cars. 
The proposed solution requires quadratic time.

We have positively tested our tool on a model of the parking space of the Federico II Hospital in Naples, one of the biggest hospitals in the South of Italy. The construction of the hospital and the annexed parking space goes back to early Sixties. Since there, no parking policy have been ever adopted: excepts for few reserved slots, a car entering the area can park in any slot. This reflects in a serious traffic congestion and an inefficient use of the slots everyday. Conversely, our approach provides, for the first time, a valid and promising solution.

In order to put it in practice, we are currently working on a mobile client application to support the drivers along the parking task, from the assignment of the slot while approaching the gate, up to the moment they leave the car. 

Our solution is the fruit of a deep analysis of the most parking-congestion-affected sites in our city, together with our strategic reasoning background. Despite being amendable, the provided solution sets the stage for future essential improvements not only of health care services offered by the hospital under exam, but also of facilities from different contexts with similar problems. Simulation results show that our solution improves notably the slot assigning with respect to the greedy parking behavior in which each car is free to select a slot according only to its own preferences. The simulation also shows that our tool is scalable and can handle very huge numbers of slots and cars.

\end{document}